\documentclass[12pt]{article}
\newtheorem{theorem}{Theorem}

\renewcommand{\baselinestretch}{1.4}

\newcommand{\filtration}{{\mbox{\boldmath$\mathcal{F}_{t-1}$}}}

\usepackage{subfigure,float}
\usepackage[default]{jasa_harvard}    

\usepackage{amsmath,amssymb}
\usepackage{color}
\usepackage{latexsym}
\usepackage{graphicx,graphics}
\DeclareGraphicsExtensions{.bmp,.jpg,.eps,.pdf}
\usepackage{float,verbatim}
\begin{document}

\title{Mean-correction and Higher Order Moments for a Stochastic Volatility Model with Correlated Errors}

\renewcommand{\baselinestretch}{1.0}
\author{Sujay Mukhoti and Pritam Ranjan\\[0.1in]
Operations Management and Quantitative Techniques,\\
Indian Institute of Management Indore, M.P., India, 453556\\
(sujaym@iimidr.ac.in, pritamr@iimidr.ac.in)
}
\date{}

\maketitle

\begin{abstract}
In an efficient stock market, the log-returns and their time-dependent variances are often jointly modelled by  stochastic volatility models (SVMs). Many SVMs assume that errors in log-return and latent volatility process are uncorrelated, which is unrealistic. It turns out that if a non-zero correlation is included in the SVM (e.g., \citeasnoun{Shephard05}), then the expected log-return at time $t$ conditional on the past returns is non-zero, which is not a desirable feature of an efficient stock market. In this paper, we propose a mean-correction for such an SVM for discrete-time returns with non-zero correlation. We also find closed form analytical expressions for higher moments of log-return and its lead-lag correlations with the volatility process. We compare the performance of the proposed and classical SVMs on S\&P 500 index returns obtained from NYSE. \\

\end{abstract}
\renewcommand{\baselinestretch}{1.5}

\section{Introduction}\label{sec:intro}

Over the last few decades different aspects of stock price movements in discrete time have been the focus of numerous research avenues. Suppose $P_t$ denotes the price of a stock at time $t$, then the continuously compounded return or log-return (here onwards referred to as return) of the stock is defined as $r_t = \log(P_t / P_{t-1})$. A stock market is said to be efficient if the price of a stock contains every available information about it. In such a market the risk involved in investing on a stock is measured by the standard deviation of $r_t$, often termed as the \emph{volatility} of the stock in finance literature. It has been noted that volatility varies over time \cite{Engle1982}. Stochastic Volatility Models (SVMs) is a popular class of models for describing the time-varying volatility of stock returns \cite{Shephard05}.

Although there are a plethora of SVMs for describing the stock returns, one of the simplest yet most popular discrete-time SVM is given by \citeasnoun{Taylor1982}, where the return process $r_t$ is a non-linear product of two independent stochastic processes, viz. an i.i.d. error process $\epsilon_t$, and a latent volatility process $h_t$, which is further modelled as an $AR(1)$. That is,
\begin{eqnarray}
  r_t & = & \exp\left\{\frac{h_{t}}{2}\right\}\epsilon_t \nonumber \\
  h_t & = & \alpha +\phi(h_{t-1}-\alpha)+\sigma \eta_t, \; \forall t=1, 2, \ldots ,
  \label{eq:SVStd}
\end{eqnarray} 
where $\alpha = E(h_t)$ is the long-range volatility, $\phi$ is the stationarity parameter, $\sigma$ measures the variability of the volatility process $h_t$, and $\epsilon_t$ and $\eta_t$ are uncorrelated i.i.d. $N(0, 1)$ errors. Hereafter this model will be referred as $SVM_0$. 


As in (\ref{eq:SVStd}), many of the new generation SVMs which are being used in the finance literature assume that $\epsilon_t$ and $\eta_t$ are independent $N(0,1)$ errors. In reality, however, $\epsilon_t$ and $\eta_t$ are often correlated \cite{Harvey1999}. Though discrete-time SVMs with non-zero $corr(\epsilon_t,\eta_t)$ have been developed earlier and are being used, they assume that $h_{t+1}$ (instead of $h_t$ as in (\ref{eq:SVStd})) depends on $\eta_t$ via AR(1) (see \emph{e.g.} \citeasnoun{Meyer2000198} and \citeasnoun{Berg2004107}). In this paper, we focus on the SVM presented in (\ref{eq:SVStd}) with correlated errors (denoted as $SVM_{\rho}$). That is, the additional assumption in (\ref{eq:SVStd}) is $corr(\epsilon_t, \eta_t) = \rho$.

It turns out that introducing a non-zero correlation between $\eta_t$ and $\epsilon_t$ in (\ref{eq:SVStd}) has an adverse effect on the admissibility of the SVM from an efficient market's viewpoint. In particular, the conditional expectation of $r_t$ given the past data, $E[r_t \mid \filtration ]$, is not zero,
where $\mathcal{F}_{t-1}$ is the space ($\sigma$-field) generated with $r_1, ..., r_{t-1}$. This zero conditional expectation of the return is a necessary requirement for an \emph{efficient market hypothesis} (EMH) (see \citeasnoun{Yu2005165} for a review).

In this paper, we propose a mean-correction for $SVM_{\rho}$ - model (\ref{eq:SVStd}) with correlated errors, such that $ E[r_t \mid \filtration ]$ becomes zero and the corrected SVM would satisfy EMH. The proposed mean-corrected model is denoted by $SVM_{\rho\mu}$. Further, \citeasnoun{Black1976167} mentioned that, usually, the amount of increment in volatility due to price fall is larger than the magnitude of reduction in the volatility due to price increase. In turn, this indicates the volatility of positive returns, $var(r_t|r_t>0)$, is less than the volatility of the negative returns, $var(r_t|r_t<0)$ resulting in skewness in return distribution. Moreover, the kurtosis quantifies the proportion of extreme values, that occur during crashes, explained by the model. We find the closed form expressions for the higher-order moments and the lead-lag correlation of the underlying return process. These descriptive statistics indicate the influence of past/future volatility on today's return.

The remainder of the article is organized as follows. Section 2 presents the main results: $SVM_{\rho\mu}$ - the mean-corrected SVM with non-zero correlation that satisfies EMH, and the closed form analytical expressions for the higher order moments and lead-lag correlation for the proposed model. For the returns of S\&P 500 NSYE, Section~3 presents a comparison between the standard zero correlation model (\ref{eq:SVStd}) and the ones with non-zero correlation. Finally Section~4 outlines the concluding remarks and a few possible future directions.

\section{Main Results}

For this section, we assume that the error terms $\epsilon_t$ and $\eta_t$ in (\ref{eq:SVStd}) have not only a constant correlation $\rho$ and i.i.d. $N(0,1)$ marginals, but they also follow a bivariate normal distribution. The proposed mean-corrected model ($SVM_{\rho\mu}$) contains an additional term $\mu$, i.e.,
\begin{eqnarray}
  r_t & = & \mu + \exp\left\{\frac{h_{t}}{2}\right\}\epsilon_t \nonumber \\
  h_t & = & \alpha +\phi(h_{t-1}-\alpha)+\sigma \eta_t, \; \forall t=1, 2, \ldots T.
  \label{eq:SVnew}
\end{eqnarray} 

Theorem~1 establishes the value of $\mu$ for which the proposed mean-corrected model (\ref{eq:SVnew}) gives zero conditional expectation $ E[r_t \mid \filtration ]$ and hence satisfy EMH. Later in this section, we derive closed form expressions for the higher-order moments, i.e., variance, skewness, and kurtosis of $r_t$, and lead-lag correlations between $r_t$ and $h_{t\pm k}$.

\begin{theorem}
For $SVM_{\rho\mu}$ in (\ref{eq:SVnew}) with $|\phi|\le 1, \sigma>0$ and $-\infty< \alpha <\infty$, if $(\epsilon_t,  \eta_t)$ follows a standard bivariate normal distribution with correlation $\rho$, the mean term
 \begin{equation}
	\mu = - \frac{\rho\sigma}{2}\exp\left\{\frac{\alpha}{2}+\frac{\sigma^2 }{8(1-\phi^2)}\right\} \label{eq:muJPR}
	\end{equation}
gives $E[r_t \mid \filtration ]=0$ and vice-versa.
\end{theorem} 

\noindent\textbf{Proof} The conditional expected return $E[r_t \mid \filtration ] =0$ gives
\begin{eqnarray}
  -\mu & = & E\left[\exp\left\{\frac{h_t}{2}\right\}\epsilon_t\right] = E\left[ \exp\left\{\frac{\alpha +\phi(h_{t-1}-\alpha)+\sigma\eta_t}{2}\right\} \epsilon_t \right] \nonumber \\
  & = & \exp\left\{\frac{\alpha}{2}\right\}\times E\left[ \exp\left\{\frac{\phi\sigma}{2}\sum_{j=1}^\infty \phi^{j-1}\eta_{t-j}\right\} \right]\times E\left[ \exp\left\{\frac{\sigma\eta_t}{2}\right\}\epsilon_t \right]. \label{eqJPR1}  
\end{eqnarray}

Since $(\epsilon_t, \eta_t)$ follows a standard bivariate normal with correlation $\rho$, the condition distribution of $\epsilon_t\mid\eta_t$ is given by $N \left( \rho\eta_t, 1-\rho^2\right)$. This conditional normal distribution and the moment generating function (mgf) of a normal distribution simplifies the third term in (\ref{eqJPR1}) as
\begin{equation}
E\left[ \exp\left\{\frac{\sigma\eta_t}{2}\right\}\epsilon_t \right]  =  E_{\eta_t}\left[\exp\left\{\frac{\sigma\eta_t}{2}\right\} \rho\eta_t \right] = \frac{\rho\sigma}{2}\exp\left\{\frac{\sigma^2}{8}\right\},\label{eqJPR2}
\end{equation}
and the second term to
\begin{equation}
%
\prod_{j=1}^{\infty}E\left[\exp\left\{\frac{\sigma \phi^j}{2}  \eta_{t-j}\right\}\right] = \exp\left\{\frac{\sigma^2}{8}\sum_{j=1}^{\infty}\phi^{2j}\right\} = \exp\left\{\frac{\sigma^2\phi^2}{8(1-\phi^2)}\right\}.
\label{eqJPR3}
\end{equation}
Hence the final expression for $\mu$ follows from (\ref{eqJPR1})-(\ref{eqJPR3}).  $\hspace{2.5in}\Box$\\

\citeasnoun{Yu2005165} tried to compute  $ E[r_t \mid \filtration ]$, but the final expression appears to be incorrect. Note that the proposed mean-correction (in Theorem~2.1) makes the model (\ref{eq:SVnew}) usable in the stock market, as it now satisfies EMH (in particular, $ E[r_t \mid \filtration ]=0$). Further, the proof of the above theorem prohibits the usage of heavy-tail distributions (like $t$ distribution) as the volatility error distribution \cite{Wang2011852} as its moment generating function would not exist resulting in in-existence of expected returns. In Section~3, we discuss the usage of this model for the index returns of S\&P500 index of New York Stock Exchange (NYSE) observed during $1^{st}$ April, 2002 - $30^{th}$ March, 2006.

\subsection{Higher-order moments}

For additional key features on the distribution of returns, we estimate higher order moments, in particular, variance, skewness and kurtosis conditional on $\filtration$. 

\begin{theorem}
For $SVM_{\rho\mu}$ in (\ref{eq:SVnew}), if Theorem~2.1 holds, then the variance of returns conditional on $\filtration$ is given by
 \begin{equation}
    V(r_t\mid \filtration)  =  \exp\left\{\alpha +\frac{\sigma^2 }{2(1-\phi^2)}\right\}\left(1+\rho^2\sigma^2-\frac{\rho^2\sigma^2}{4} \exp\left\{-\frac{\sigma^2 }{4(1-\phi^2)}\right\} \right). \label{eqSVLMVar}
 \end{equation}
\end{theorem}

\noindent\textbf{Proof} Following the definition of variance, 
  \begin{eqnarray*}
    V(r_t\mid \filtration) &=& E[r_t^2\mid \filtration] - 0^2 \\
    & = & E\left[ \exp\{h_t\}\epsilon_t^2 \right]-\mu^2 \\
  & = & \exp\left\{\alpha\right\}\times E\left[ \exp\left\{\sigma\sum_{j=1}^\infty \phi^{j}\eta_{t-j}\right\} \right]\times E\left[ \exp\left\{\sigma\eta_t\right\}\epsilon_t^2 \right] - \mu^2\\
    & = & \exp\left\{\alpha + \frac{\sigma^2\phi^2}{2(1-\phi^2)}\right\} (1 + \rho^2\sigma^2) \exp\left\{\frac{\sigma^2}{2}\right\} - \mu^2 \quad \textrm{(as in (\ref{eqJPR1})-(\ref{eqJPR3}))}.
  \end{eqnarray*}
The final result follows by substituting the value of $\mu$ from Theorem~2.1. $\hspace{1in}\Box$\\

The expressions of the conditional mean and variance are the most crucial components in finding the skewness and kurtosis statistics. For $SVM_{\rho\mu}$ in (\ref{eq:SVnew}), under the same conditions as in Theorem~2.2, the skewness conditional on $\filtration$ is measured by $\mu_3/(Var(r_t\mid\filtration))^{3/2}$, where
\begin{eqnarray} \label{eqSVMmu3}
    \mu_3  &=&  \nonumber
	\frac{3\rho\sigma}{2} exp\left\{\frac{3\alpha}{2}+\frac{9\sigma^2 }{8(1-\phi^2)}\right\} \left[ 3+ \frac{9\sigma^2\rho^2}{4}+\frac{\rho^2\sigma^2}{6} exp\left\{-\frac{3\sigma^2 }{4(1-\phi^2)}\right\} \right.\\
	& &\left. -\left(1+\rho^2\sigma^2 \right)exp\left\{-\frac{\sigma^2 }{2(1-\phi^2)}\right\} \right]. 
%
	%
\end{eqnarray}	

The proof of (\ref{eqSVMmu3}) starts with $\mu_3 = E[r_t^3\mid \filtration]$, and proceeds in the exact same manner as in Theorems~2.1 and 2.2. Similarly the closed form expression of kurtosis can also be found as $\mu_4/(Var(r_t\mid\filtration))^2$, where
\begin{eqnarray} \label{eqSVMmu4}
    \mu_{4}  &=&  \nonumber
	\exp\left\{2\alpha+\frac{2\sigma^2}{(1-\phi^2)} \right\} \times \left[ \frac{3}{2} \rho^2\sigma^2(1+\sigma^2\rho^2)\exp\left\{\frac{-5\sigma^2}{4(1-\phi^2)}\right\} \right. \\
	&& +  \left( 3 + 24 \rho^2\sigma^2+16\rho^4\sigma^4\right) - \frac{3}{16}\rho^4\sigma^4 \exp\left\{\frac{-3}{2}\frac{\sigma^2}{(1-\phi^2)}\right\} \nonumber \\
 && \left. -9\rho^2\sigma^2  \left( 1+\frac{3}{4}\rho^2 \sigma^2\right) \exp\left\{\frac{-3\sigma^2}{4(1-\phi^2)}\right\} \right].
\end{eqnarray}	

As expected all four descriptive statistics found here depends heavily  on $corr(\epsilon_t, \eta_t) = \rho$. On a closer inspection of these statistics, we see that $\rho=0$ (i.e., the classical SVM by \cite{Taylor1982}) gives $\mu=0, \mu_3=0$, 
$$Var(r_t\mid\filtration)= \exp\left\{\alpha +\frac{\sigma^2 }{2(1-\phi^2)}\right\} \quad \textrm{and} \quad \mu_4 = 3\exp\left\{2\alpha+\frac{2\sigma^2}{(1-\phi^2)} \right\}.
$$

Note that the simplified expressions found here are consistent with the ones reported by \citeasnoun{Ghysels1996119}, and hence the proposed model $SVM_{\rho\mu}$ (\ref{eq:SVnew}) is a generalization of the classical model $SVM_0$ in (\ref{eq:SVStd}). Next we investigate the conditional (on $\filtration$) dependence between the current returns and past, current and future  volatility. 

\subsection{Lead-lag correlations}

In this section, we wish to estimate three quantities: (1) dependence between the current returns and current volatility, $corr(r_t, h_t|\filtration)$, (2) the potential influence of current returns on future volatility, $corr(r_t, h_{t+k}|\filtration)$, and (c) the influence of past volatility on current returns, $corr(r_t, h_{t-k}|\filtration)$. Though empirical estimation of such quantities is not uncommon, e.g., in \citeasnoun{Bollerslev2006}, our aim is to find closed analytical expression for these descriptive measures under $SVM_{\rho\mu}$ specification.

Since $var(r_t| \filtration)$ is given by (\ref{eqSVLMVar}) and $var(h_t|\filtration) = \sigma^2/(1-\phi^2)$, we only need to find the expressions for the conditional covariances. First, we recall that under the proposed model, the conditional means are $E(r_t|\filtration) = 0$ and $E(h_t|\filtration) = \alpha$. Now, if we assume that $corr(r_t, h_t|\filtration)=\sigma_{rh}$, then 
\begin{eqnarray*}
cov(r_t, h_{t+1}) = E[r_t(h_t-\alpha)] = E[r_t(\phi(h_{t}-\alpha) + \sigma\eta_{t+1})] = \phi E[r_t(h_t-\alpha)] = \phi \sigma_{rh},
\end{eqnarray*}
which further implies that $cov(r_t, h_{t+k}) = \phi^k\sigma_{rh}$ for $k \ge 1$. By applying the key mathematical techniques (i.e., properties of expectation, normal mgf and the expansion of $h_t = \alpha + \sigma \sum_{j=1}^{\infty}\eta_{t-j}\phi^j$) used in proving results of Section~2.1, one can easily show that 
\begin{eqnarray*}
\sigma_{rh} = cov(r_t, h_t|\filtration) &=& \rho\sigma\exp\left\{\frac{\alpha}{2} + \frac{\sigma^2}{8(1-\phi^2)}\right\} \times \left\{1 + \frac{\sigma^2}{4(1-\phi^2)}\right\},\\
cov(r_t, h_{t-k}|\filtration) &=& \sigma_{rh} \cdot \phi^k \cdot \left[\frac{\sigma^2}{4(1-\phi^2)}\right]\big/\left[1+\frac{\sigma^2}{4(1-\phi^2)}\right].
\end{eqnarray*}

Clearly, both the lead ($cov(r_t, h_{t+k})$) and lag ($cov(r_t, h_{t-k})$) covariances are smaller than the contemporaneous covariance $cov(r_t, h_{t-k}|\filtration)$. The contemporaneous correlation can be interpreted as \emph{feedback effect} of volatility change on future returns, whereas the impact of return change on future volatility is termed as \emph{leverage effect}. \citeasnoun{Wu2000} found that volatility feedback effect is stronger than leverage effect. The closed form expressions we have derived above provide a theoretical proof of the mentioned findings under $SVM_{\rho\mu}$ specification. Moreover, \citeasnoun{Bollerslev2006} have empirically observed that the lag-correlation with lag $h$ is smaller than lead correlation with lead $h$ which we have established theoretically. Further note that all these covariances and hence correlations vanish if $\rho=corr(\epsilon_t, \eta_t) = 0$. Next, we compare the goodness of fit of the three stochastic volatility models, $SVM_0$ (classical - with zero correlation), $SVM_{\rho}$ (with correlation $\rho$) and $SVM_{\rho\mu}$ (mean-corrected with correlation $\rho$), for a real data on returns.

\section{Example: S\&P 500 NYSE}

\begin{figure}[!ht]
\centering
	\includegraphics[width=5.5in]{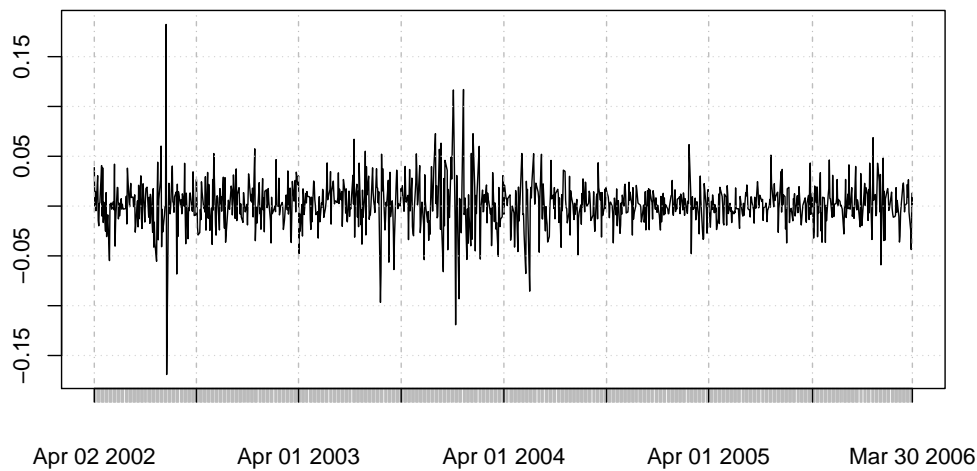}
	\caption{Time plot of S\&P500 returns during April 01, 2002 -- March 30, 2006}
	\label{fig:TimePlotOfSP500Returns}
\end{figure}

In this paper, we compare the performance of the three models ($SVM_0, SVM_{\rho}, SVM_{\rho\mu}$) on the index returns of Standard and Poor 500 index (S\&P500) obtained from New York Stock Exchange during April 01, 2002 -- March 30, 2006. We selected this period to avoid extreme behaviour during ``2000 -- 2002 dot-com bubble" and ``2008 Lehman Brothers' crash". Figure~\ref{fig:TimePlotOfSP500Returns} displays the time-plot of the returns of 1008 trading days (less than the total number of calendar days).

From Figure~\ref{fig:TimePlotOfSP500Returns} one can infer that the volatility is relatively high during September 2003 and June 2004, whereas during October 2004 to April 2005, the volatility is relatively lower than usual. A few descriptive statistics of the observed returns are as follows:
\begin{table}[h!]\centering
\begin{tabular}{ll}
mean = 0.0014, & variance = 0.0005,\\
skewness =  0.0329,& kurtosis = 10.9813.\\
\end{tabular}
\end{table}

We follow \citeasnoun{Meyer2000198}, and use the same Markov Chain Monte Carlo (MCMC) algorithm implemented in Just Another Gibbs Sampler (JAGS) for fitting the classical model $SVM_0$. For fitting the other two models, $SVM_{\rho}$ and $SVM_{\rho\mu}$, we slightly modify the JAGS code to include the $corr(\epsilon_t, \eta_t)=\rho$ and $\mu$ (derived in Theorem~2.1). For implementing $SVM_0$ in JAGS, the hierarchical model structure is characterized by 
\[ r_t\mid (h_t,h_{t-1},\ldots h_1, h_0; \alpha,\phi,\sigma) \sim N\left(0, \exp\left\{ h_t \right\}\right),\]
\[ \textrm{and} \quad h_t\mid (h_{t-1},\ldots h_1, h_0; \alpha,\phi,\sigma) \sim N\left(\alpha+\phi(h_{t-1}-\alpha), \sigma^2\right). \]

For $SVM_{\rho}$, the mean and variance of the conditional distribution of $r_t$ changes to
\[ r_t\mid (h_t,\ldots, h_0; \alpha,\phi,\sigma) \sim N\left(\frac{\rho \ e^{h_t/2}}{\sigma}(h_t-\alpha-\phi(h_{t-1}-\alpha)), e^{h_t}(1-\rho^2)\right),\]
and the conditional distribution of $h_t$ remains the same. Similarly, the implementation of the mean-corrected model $SVM_{\rho\mu}$ is characterized by updating the mean and variance of the conditional distribution of $r_t$ to
\[ r_t\mid (h_t,\ldots, h_0; \alpha,\phi,\sigma) \sim N\left(\mu+\frac{\rho\ e^{h_t/2}}{\sigma}(h_t-\alpha-\phi(h_{t-1}-\alpha)), e^{h_t}(1-\rho^2)\right).\]

\begin{figure}[!h]\centering
	\subfigure[$\alpha$ (expected volatility)]{\includegraphics[width=3.1in]{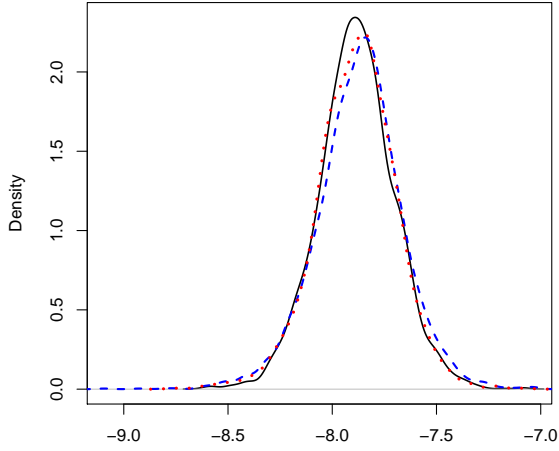}}
	\subfigure[$\phi$ (stationarity parameter)]{\includegraphics[width=3.1in]{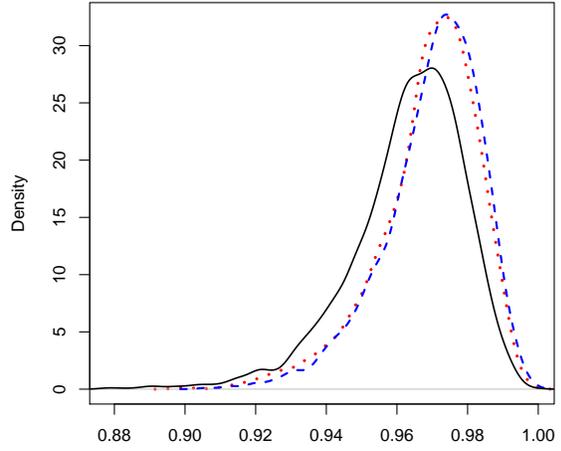}}

	\vspace{-0.4in}
	
	\subfigure[$\sigma$ (variability in volatility)]{\includegraphics[width=3.1in]{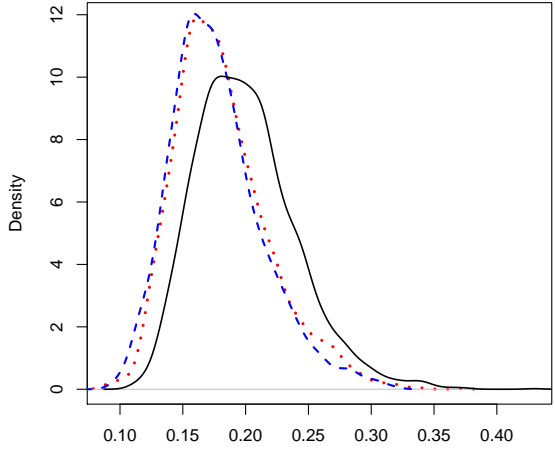}}
	\subfigure[$\rho = corr(\epsilon_t, \eta_t)$]{\includegraphics[width=3.1in]{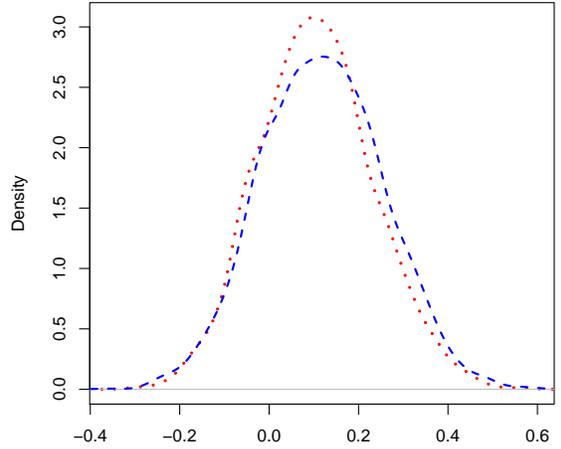}}
	\caption{Posterior distribution of $\Theta$ for the three models. The black solid curves represent $SVM_0$, blue dashed curves are from $SVM_{\rho}$ model, and the red dotted curves are obtained from the proposed model $SVM_{\rho\mu}$.}
	\label{fig:boxplots}
\end{figure}

The parameters of interest are $(\alpha, \phi, \rho, \sigma)= \Theta$ (say). We use the same prior (including the hyperparameters) for $\alpha, \phi$ and $\sigma$ as in \citeasnoun{Meyer2000198}, and a non-informative $Unif(-1,1)$ prior for the correlation parameter. The posterior of $\Theta$ and $\mathcal{H}=\{h_t, h_{t-1},...\}$ given the data $\{r_t,r_{t-1},...\}$ is obtained via JAGS. We set the total length of chains to be 180,000, out of which 30,000 was the burn-in, and from the remaining 150,000 posterior realizations (with the thinning of every $50^{th}$ realization) were used (i.e., 3000 realizations in total) to obtain the plug-in estimates of the parameters. The thinning process facilitates a safeguard against the chain dependency in the sampling process. Figure~\ref{fig:boxplots} shows the density plots of the posterior distribution of $\Theta$ for the three models, $SVM_0, SVM_{\rho}$ and $SVM_{\rho\mu}$. We have not included the traceplots, as all parameters converge nicely and the plots do not reveal anything extra. The plug-in estimates of the parameters are obtained via posterior mean and variance (summarized in Table~1).

\begin{table}[h!]\centering \caption{Plug-in estimators of $\Theta = (\alpha, \phi, \sigma, \rho)$ for the three models. The numbers in parentheses show the standard deviation of the posterior realizations.}
\begin{tabular}{c|lll}
Parameter & $SVM_0$ & $SVM_{\rho}$ & $SVM_{\rho\mu}$ \\
\hline
$\alpha$ & -7.88 & -7.87      & -7.88   \\
		 & (0.1837)& (0.2077) & (0.192)	\\
$\phi$   & 0.96  & 0.97 & 0.96 \\
		 & (0.016)& (0.014)& (0.014) \\
$\sigma$ & 0.2 & 0.177      & 0.18 \\
		 & (0.04)& (0.034)&(0.038) \\
$\rho$   &       & 0.1185 & 0.105 \\
		 & 		 & (0.1362)& (0.1278) \\
\hline
\end{tabular} 
\end{table}

Table~1 shows that the posterior estimates of the parameters in $SVM_0,\; SVM_{\rho}$ and $SVM_{\rho\mu}$ are similar. Further, the near-unity estimate of $\phi$ indicates presence of strong volatility clustering. The estimate of the correlation parameter $\rho$ is small yet positive, which is similar to the findings of \citeasnoun{Schwert1987} and \citeasnoun{campbell1992}. This may be taken as an indication of no significant effect of current return on future volatility.  

Figure~\ref{fig:boxplots} shows that the posterior distributions of the parameters for $SVM_0$, $SVM_\rho$ and $SVM_{\rho\mu}$ are different in their kurtosis. A general pattern that can be noticed is that posterior distributions of the parameters under $SVM_0$ are more leptokurtic compared to their counter parts under other two models except for $\alpha$. Importantly, too strong volatility clustering is more probable under $SVM_{\rho}$ and $SVM_{\rho\mu}$ compared to $SVM_0$. In case of variance of volatility, posterior distribution under $SVM_0$ indicates higher values compared to the other two models. Comparing the posterior distributions of $\rho$ under $SVM_\rho$ and $SVM_{\rho\mu}$, the former shows higher probability of being positive valued relative to the latter. 

As per Figure~2(d), $\rho$ is very small (close to zero), and thus, it is expected that the proposed model would not provide significant additional strength in modelling the returns data.

We now compare the three models using the descriptive measures (mean, variance, skewness and kurtosis), three lead-lag correlations, mean deviance over the posterior distribution, and the mean square prediction error (MSPE): $\sum_{t=1}^T \hat{r}_t^2/T$. The deviance function, suggested by \cite{Dempster1974}, is  
\[ D(\Theta) =-2\log f(r\mid \Theta, \mathcal{H})+2\log g(r), \]
where $f(r\mid \Theta, \mathcal{H})$ is the likelihood for a given realization of $\Theta$ and $\mathcal{H}$, and $g(r)$ is the normalizing constant. Table~2 presents the plug-in values of these ``goodness of fit" measures for the three models.

\begin{table}[h!]\centering \caption{Goodness of fit measures for the true data and the three models.}
\begin{tabular}{l|llll}
GOF measure & True data & $SVM_0$ & $SVM_{\rho}$ & $SVM_{\rho\mu}$ \\
\hline
Mean     & 0.0014 & 0      & --       & $-4.05 \times 10^-6$\\
Variance & 0.0005 & 0.0005 & 0.0005 & 0.0005 \\
Skewness & 0.0329 & 0      & 0.0856 & 0.0769 \\
Kurtosis & 10.981 & 5.196 & 5.105 & 5.076 \\
\hline
$corr(r_t,h_t)$ & & &0.0305 & 0.0276 \\
$corr(r_t, h_{t-10})$ & &&& 0.0053 \\
\hline
Deviance & & -5019& -5033& -5043 \\
MSPE ($\times 10^{-7}$) & & 0.178& 12.94& 9.298\\
\hline
\end{tabular} 
\end{table}

Since $\rho \approx 0.1$ (very small), the estimated mean is also small $\mu = -4.05 \cdot 10^{-6}$. Thus all three models would behave very similarly (which is reflected in the estimated moments under the three models). Surprisingly plug-in estimates of kurtosis obtained from all three models under-estimates the kurtosis measured from the data. Deviance values indicate that $SVM_{\rho\mu}$ provides a slightly better fit compared to the other two models. On the other hand, MSPE values indicate that the basic SVM provides better prediction among the three models. Though the numerical results presented through the S\& P 500 NYSE example do not provide sufficient evidence for $SVM_{\rho\mu}$ giving additional information than $SVM_0$, it certainly is the generalization of $SVM_0$ and an example with large $\rho=corr(\epsilon_t, \eta_t)$ might have given more convincing evidence.

\section{Concluding Remarks}

In this paper, we have proposed a mean-correction for the SVM with correlation between $\epsilon_t$ and $\eta_t$. This mean-correction step enables the conditional expected return to be zero, which is a necessary condition for a good SVM (i.e., a model that adhere to the EMH). We have also found the closed form analytical expression for the higher moments of returns and lead-lag correlation between the return and volatility. 

From S\&P500 example, we see that most of the empirical observations on statistical properties of returns are reflected through all the three models. However, $SVM_{\rho\mu}$ gives a slightly better fit to the data (in terms of average deviance) compared to the classical model $SVM_0$ as well as $SVM_\rho$. A close look at this research endeavour  generates several interesting and challenging research problems.

First, the estimated error correlation $\rho$ turns out to be positive despite the fact that return and its volatility move in opposite directions \cite{Nelson1991}. \citeasnoun{GJR1993} attributed this discrepancy due to mis-specification in the underlying SVM, which is caused by not accounting for the size discrepancy in volatility change due to up or down movement of price. The authors have shown that if the size discrepancy is accounted for then $\rho$ becomes negative. This result demonstrates that $\rho$ alone can not explain the asymmetric response of return to its volatility sufficiently. As we have pointed out in the introduction that this size discrepancy can be interpreted as different conditional variances (or volatility) for positive and negative returns, which leads to skewed return distribution instead of a Gaussian one, a new model can be developed by extending $SVM_{\rho \mu}$ in the line of \citeasnoun{Abanto-Valle20102883}. 



Second, the observed kurtosis from the data is not completely explained by the model based estimates of kurtosis. Indeed, the significant difference between empirical kurtosis and the model based estimates again suggests non-normality of the return error distribution. The problem can be tackled in two ways- (1) introducing jumps in returns or (2) allowing the return error to be heavy-tailed (\emph{e.g.} Student's $t$). Notice, adding a jump to the return only explains transient changes (as seen on 8$^{th}$ \& 9$^{th}$ August, 2002 ) and does not cause the return distribution to change permanently whereas jump in both return and volatility explains persistent effects of extreme values (\emph{e.g.} September, 2003 -- June, 2004). $SVM_{\rho \mu}$ can further be generalized by including jumps in return and volatility \cite{Eraker2003} following the 1$^{st}$ line of argument and using skew Student's-$t$ distributions following the 2$^{nd}$ line of argument \cite{Dipak2015}.

Although continuous time stochastic volatility has been studied extensively in the literature, the comparatively new discrete-time SVM brings out new interesting features such as leverage effect and feedback eefect which occurs due to lagged reaction between return and its volatility. In this paper we have established that the empirically observed pattern of leverage effect and lagged correlations \cite{Bollerslev2006} are explained by $SVM_{\rho\mu}$. In particular, we have shown that the correlation between current return and future volatility is maximum in magnitude at lead $0$ (or contemporaneously) and the future leverage effects disappear exponentially with the lead time. Indeed, strong volatility clustering effect indicates more persistent leverage effect. It may also be noted that the existing practice of assuming $h_{t+1}=\alpha+\phi(h_t-\alpha)+\sigma \eta_t$ (instead of $h_t$) and $corr(\epsilon_t,\eta_t)=\rho$ for a correct SVM specification would not support the empirical observation on contemporaneous correlation. 

Mean correction to the contemporaneously correlated SVM has a very important application. Zero conditional expected returns is also referred to as the martingale difference property, which is a necessary and sufficient condition for \emph{no arbitrage} - which further leads to the existence of option pricing kernel (\citeasnoun{Back1991}, \citeasnoun{Shephard2009}). That is, we believe that the proposed mean-correction strategy can also be used in option pricing.

\bibliographystyle{ECA_jasa}
\bibliography{SVMRef}

\end{document}